\documentstyle[12pt]{article}
\def\be{\begin{eqnarray}}
\def\ee{\end{eqnarray}}
\def\ba{\begin{array}}
\def\ea{\end{array}}
\def\p{\phi}

\def\s{\Sigma}
\def\t{\Theta}
\def\n{\nabla}

\def\pa{\partial}

\def\G{{\cal G}}
\def\B{{\cal B}}
\def\A{{\cal A}}
\def\X{{\cal X}}

\def\S{{\cal S}}

\def\L{{\cal L}}

\def\D{^{(D)}}
\def\Z{{\cal Z}}

\begin{document}
\begin{center}
{\LARGE 
{O(d+1,d+n+1)--invariant Formulation of Stationary\\
\vskip 0.3cm
Heterotic String Theory
}}
\end{center}

\vskip 1.5cm

\begin{center}
{\bf \large 
{Nandinii Barbosa--Cendejas}}
\end{center}
\begin{center}
Escuela de Ciencias F\'\i sico--Matem\'aticas,\\
Universidad Michoacana de San Nicol\'as de Hidalgo\\
e-mail: nandinii@fismat.umich.mx
\end{center}

\vskip 5mm
\begin{center}
and
\end{center}
 
\vskip 3mm
\begin{center}
{\bf \large {Alfredo Herrera--Aguilar}}
\end{center}
 
\begin{center}
Instituto de F\'\i sica y Matem\'aticas\\
Universidad Michoacana de San Nicol\'as de Hidalgo\\
Apdo. Postal 2-82, Morelia, Mich., M\'exico\\
e-mail: herrera@zeus.umich.mx
\end{center}
 
%%%%%%%%%%%%%%%%%%%%%%%%%%%%%%%%%%%%%%%%%%%%%%%%%%%%%%%%%%%%%%%%%%%%%%%%%%%%%%
%%%%%%%%%%%%%%%%%%%%%%%%%%%%%%%%%%%%%%%%%%%%%%%%%%%%%%%%%%%%%%%%%%%%%%%%%%%%%%
\begin{abstract} 
We present a pair of symmetric formulations of the matter sector of
the stationary effective action of heterotic string theory 
that arises after the toroidal compactification of $d$ dimensions. 
The first formulation is written in terms of a pair of matrix  
potentials $Z_1$ and $Z_2$ which exhibits a clear symmetry between 
them and can be used to generate new families of solutions on the
basis of either $Z_1$ or $Z_2$; the second one is an 
$O(d+1,d+n+1)$--invariant formulation which is written in terms of 
a matrix vector ${\bf W}$ endowed with an $O(d+1,d+n+1)$--invariant 
scalar product which linearizes the action of the $O(d+1,d+n+1)$
symmetry group on the coset space $O(d+1,d+n+1)/[O(d+1)\times O(d+n+1)]$;
this fact opens as well a simple solution--generating technique 
which can be applied on the basis of known solutions. A special class 
of extremal solutions is indicated by asuming a simple ansatz for the
matrix vector ${\bf W}$ that reduces the equation of motion to the Laplace 
equation for a real scalar function.
\end{abstract}
%%%%%%%%%%%%%%%%%%%%%%%%%%%%%%%%%%%%%%%%%%%%%%%%%%%%%%%%%%%%%%%%%%%%%%%%%%%%%%
%%%%%%%%%%%%%%%%%%%%%%%%%%%%%%%%%%%%%%%%%%%%%%%%%%%%%%%%%%%%%%%%%%%%%%%%%%%%%%
%\pacs{ \noindent PACS numbers: 04.20.Jb, 04.50.+h %} \newpage
%%%%%%%%%%%%%%%%%%%%%%%%%%%%%%%%%%%%%%%%%%%%%%%%%%%%%%%%%%%%%%%%%%%%%%%%%%%%%%
%%%%%%%%%%%%%%%%%%%%%%%%%%%%%%%%%%%%%%%%%%%%%%%%%%%%%%%%%%%%%%%%%%%%%%%%%%%%%%
\section{Introduction}
%%%%%%%%%%%%%%%%%%%%%%%%%%%%%%%%%%%%%%%%%%%%%%%%%%%%%%%%%%%%%%%%%%%%%%%%%%%%%%
%%%%%%%%%%%%%%%%%%%%%%%%%%%%%%%%%%%%%%%%%%%%%%%%%%%%%%%%%%%%%%%%%%%%%%%%%%%%%% 
At low energies the heterotic string theory leads to an effective field  
theory of massless fields which describes supergravity coupled to some 
matter fields. In \cite{marsch}--\cite{sen434} it was shown that when 
considering the toroidal compactification from $D$ to $3$ dimensions using a 
Kaluza--Klein ansatz, the resulting theory turns out to be a nonlinear 
$\sigma$--model with values in the symmetric coset space 
$SO(d+1,d+n+1)/S[O(d+1)\times O(d+n+1)]$. It is well known that the theory of 
symmetric spaces provides a convenient framework for discussing and understanding 
the internal symmetries of the $\sigma$--models. Later on, an alternative 
representation of this effective field theory in terms of a couple of matrix 
Ernst potentials (MEP) was proposed in \cite{hk3}. This formulation is, in fact, 
a matrix 
generalization of the nonlinear $\sigma$--model parametrization of the stationary
Einstein--Maxwell (EM) theory and enables us to extrapolate the results obtained 
in the EM theory to the heterotic string realm; we review this formalism in 
Section 2. Further development in this direction was achieved in \cite{hk5}--\cite{ok1}
by introducing a single rectangular matrix potential which transforms linearly 
under the action of the charging symmtery group. Several families of solutions 
have been constructed making use of the analogy between both theories (see, for
instance, \cite{ky}--\cite{aha1}), however, the full charge parametrized stationary 
axisymmetric black hole solution is still missing \cite{youm}. Most of the 
constructed solutions were obtained by making use of solution generating 
techniques based on the symmetries of the effective theory and their equations 
of motion.

In the framework of this motivation we present in Section 3 a particularly 
symmetric formulation of the matter sector of the effective field
theory of heterotic string in terms of a pair of matrix potentials that enter 
the effective action in a completely symmetric way. This fact allows us to 
construct new solutions for the hole theory starting from a solution for the 
truncated theory in terms of one of the matrix potentials. We discuss as well the 
restrictions under which this procedure can consistently take place. Subsequently
we give an explicit $O(d+1,d+n+1)$--invariant formulation of the effective theory
in terms of a matrix vector endowed with an $O(d+1,d+n+1)$--invariant scalar 
product. This formulation linearizes the action of the $O(d+1,d+n+1)$ symmetry 
group on the coset space $O(d+1,d+n+1)/O[(d+1)\times (d+n+1)$ and open the 
possibility of applying a solution--generating technique in order to construct 
new families of solutions. We conclude this Section by choosing a simple ansatz
for the matrix vector which reduces the equation of motion to the Laplace 
equation for a real scalar function and yields a family of extremal solutions. 
Finally, in Section 4 we present our conclusions and discuss on the further 
development and applications of this formalism.

%%%%%%%%%%%%%%%%%%%%%%%%%%%%%%%%%%%%%%%%%%%%%%%%%%%%%%%%%%%%%%%%%%%%%%%%%%%%%%
%%%%%%%%%%%%%%%%%%%%%%%%%%%%%%%%%%%%%%%%%%%%%%%%%%%%%%%%%%%%%%%%%%%%%%%%%%%%%%
\section{The effective action and matrix Ernst potentials}
%%%%%%%%%%%%%%%%%%%%%%%%%%%%%%%%%%%%%%%%%%%%%%%%%%%%%%%%%%%%%%%%%%%%%%%%%%%%%%
%%%%%%%%%%%%%%%%%%%%%%%%%%%%%%%%%%%%%%%%%%%%%%%%%%%%%%%%%%%%%%%%%%%%%%%%%%%%%%
Let us consider the effective action of the heterotic string at tree level
\be
\S\D\!=\!\int\!d\D\!x\!\mid\!
G\D\!\mid^{\frac{1}{2}}\!e^{-\p\D}\!(R\D\!+\!
\p\D_{;M}\!\p^{(D);M}
\!-\!\frac{1}{12}\!H\D_{MNP}H^{(D)MNP}\!-\!
\frac{1}{4}F^{(D)I}_{MN}\!F^{(D)IMN}),
\ee
where
\be
F^{(D)I}_{MN}\!=\!\pa_MA^{(D)I}_N\!-\!\pa _NA^{(D)I}_M, \quad
H\D_{MNP}\!=\!\pa_MB\D_{NP}\!-\!\frac{1}{2}A^{(D)I}_M\,F^{(D)I}_{NP}\!+\!
\mbox{{\rm cycl. perms. of} \, M,\,N,\,P.}
\nonumber
\ee
Here $G\D_{MN}$ is the metric, $B\D_{MN}$ is the
anti--symmetric Kalb-Ramond field, $\p\D$ is the dilaton and 
$A^{(D)I}_M$
is a set of Abelian $U(1)$ vector fields ($I=1,\,2,\,...,n$). In the 
consistent critical case $D=10$ and $n=16$, but we shall leave these 
parameters arbitrary for the sake of generality since when $d=1$ and 
$n=6$ the matter content of the considered effective field theory
corresponds to that of $D=N=4$ supergravity, and when $d=n=1$, to
that of Einstein--Maxwell Dilaton--Axion theory; moreover, several cases 
have been considered in the literature using different values of
$d$ and $n$.
 
In \cite{ms}--\cite{sen434}
it was shown that after the compactification of this model on a
$D-3=d$--torus, the resulting stationary theory possesses the
$SO(d+1,d+1+n)$ symmetry group and describes gravity coupled to the 
following set of three--dimensional fields:
 
\noindent a) scalar fields
\be
G\!\equiv\!G_{mn}\!=
\!G\D_{m+3,n+3},\,\,\,
B\!\equiv\!B_{mn}\!=
\!B\D_{m+3,n+3},\,\,\,
A\!\equiv\!A^I_m\!=
\!A^{(D)I}_{m+3},\,\,\,
\p\!=\!\p\D\!-\!\frac{1}{2}{\rm ln|det}\,G|,
\ee
\noindent b)tensor fields
\be
g_{\mu\nu}\!=\!e^{-2\p}\!\left(G\D_{\mu\nu}\!-
\!G\D_{m+3,\mu}G\D_{n+3,\nu}G^{mn}\right),\,
B_{\mu\nu}\!=\!B\D_{\mu\nu}\!\!-\!4B_{mn}A^m_{\mu}A^n_{\nu}\!-\!
2\!\left(A^m_{\mu}A^{m+d}_{\nu}\!-\!A^m_{\nu}A^{m+d}_{\mu}\right),
\ee 
\noindent c)vector fields $A^{(a)}_{\mu}=
\left((A_1)^m_{\mu},(A_2)^{m+d}_{\mu},(A_3)^{2d+I}_{\mu}\right)$
\be
(A_1)^m_{\mu}\!=\!\frac{1}{2}G^{mn}G\D_{n+3,\mu},\,
(A_3)^{I+2d}_{\mu}\!=\!-\frac{1}{2}A^{(D)I}_{\mu}\!+\!A^I_nA^n_{\mu},\,
(A_2)^{m+d}_{\mu}\!=\!\frac{1}{2}B\D_{m+3,\mu}\!\!-\!B_{mn}A^n_{\mu}\!+\!
\frac{1}{2}A^I_{m}A^{I+2d}_{\mu}
\ee                                                         
where the subscripts $m,n=1,2,...,d$; and $a=1,...,2d+n$. Following 
\cite{sen434}, in this paper 
we set $B_{\mu\nu}=0$ in order to remove the effective 
cosmological constant from our consideration.

In three dimensions all vector fields can be dualized on--shell as follows
\begin{eqnarray}
\nabla\times\overrightarrow{A_1}&=&\frac{1}{2}e^{2\p}G^{-1}
\left(\nabla u+(B+\frac{1}{2}AA^T)\nabla v+A\nabla s\right),
\nonumber                          \\
\nabla\times\overrightarrow{A_3}&=&\frac{1}{2}e^{2\p}
(\nabla s+A^T\nabla v)+A^T\nabla\times\overrightarrow{A_1},\\
\nabla\times\overrightarrow{A_2}&=&\frac{1}{2}e^{2\p}G\nabla v-
(B+\frac{1}{2}AA^T)\nabla\times\overrightarrow{A_1}+
A\nabla\times\overrightarrow{A_3}.
\nonumber                            
\end{eqnarray}
Thus, the effective stationary theory describes gravity $g_{\mu\nu}$
coupled to the scalars $G$, $B$, $A$, $\p$ and pseudoscalars 
$u$, $v$, $s$. All matter fields can be arranged in the following pair 
of MEP
\be
\X=
\left(
\ba{cc}
-e^{-2\p}+v^TXv+v^TAs+\frac{1}{2}s^Ts&v^TX-u^T \cr
Xv+u+As&X
\ea
\right), \quad \qquad
\A=\left(
\ba{c}
s^T+v^TA \cr
A
\ea
\right),
\ee
where $X=G+B+\frac{1}{2}AA^T$. These matrices have dimensions
$(d+1) \times (d+1)$ and $(d+1) \times n$, respectively. 
Some words about the physical meaning of of the fields are in order.
The relevant information of the gravitational field is encoded in 
$X$, whereas its rotational nature is hidden in $u$; $\p$ is the dilaton
field, $v$ is related to multidimensional components of the Kalb--Ramond
field, $A$ and $s$ stand for electric and magnetic potentials, respectively.

In terms of MEP, the effective three--dimensional 
theory adopts the form                                                                                 
\be
^3\S\!=
\!\int\!d^3x\!\mid g\mid^{\frac{1}{2}}\!\{\!-\!R\!+
\!{\rm Tr}[\frac{1}{4}\left(\nabla \X\!-\!\nabla \A\A^T\right)\!\G^{-1}
\!\left(\nabla \X^T\!-\!\A\nabla \A^T\right)\!\G^{-1}
\!+\!\frac{1}{2}\nabla \A^T\G^{-1}\nabla \A]\},
\ee
where $\X=\G+\B+\frac{1}{2}\A\A^T$, hence the matrices \,
$\G=\frac{1}{2}\left(\X+\X^T-\A\A^T\right)$ and
$\B=\frac{1}{2}\left(\X-\X^T\right)$ read
\be
\G=
\left(
\ba{cc}
-e^{-2\p}+v^TGv&v^TG \cr
Gv&G
\ea
\right), \quad 
\B=\left(
\ba{cc}
0&v^TB-u^T \cr
Bv+u&B
\ea
\right).
\ee
In \cite{hk3} it was shown that there exist a map between the 
stationary actions of the heterotic string and Einstein--Maxwell 
theories
\be
\X\longleftrightarrow E, \quad \A\longleftrightarrow F,
\nonumber
\ee
\be
{\it matrix\,\, transposition}\quad\longleftrightarrow\quad
{\it complex\,\, conjugation},
\ee
where $E$ and $F$ are the complex Ernst potentials of the 
Einstein--Maxwell theory \cite {e}. Thus, the map (9) allows us to generalize
the results obtained in the EM theory to the heterotic string one 
using the MEP formulism. It is worth noticing that in the right hand side we
have complex functions, whereas in the left hand side we have real matrices
(hence the transposition instead the complex conjugation) that obey the usual 
rules of matrix algebra.
%%%%%%%%%%%%%%%%%%%%%%%%%%%%%%%%%%%%%%%%%%%%%%%%%%%%%%%%%%%%%%%%%%%%%%%%%%%%%%
\section{Symmetric Formulations}

We begin this Section by reformulating the matter sector of the 
three--dimensional effective
field theory of the heterotic string (6). First we give a particularly symmetric
representation of the effective theory in terms of two potentials and then we
rewrite it in an explicit $O(d+1,d+n+1)$--invariant form.

Thus, if we substitute $\X=2(Z_1+\s)^{-1}-\s$ and $\A=\sqrt{2}(Z_1+\s)^{-1}Z_2$, 
where $Z_1$and $\s$ are matrices of dimension $d+1$,
$Z_2$ is a $(d+1)\times n$--matrix and $\s=diag(-1,-1,1,1,...1)$, 
the matter sector of the action (7) takes the form
\be
\S_{matt}=\int d^3x\mid g\mid^{\frac{1}{2}}
Tr\left[\n Z_1\left(\s+\s Z_1^T\t^{-1}Z_1\s\right)\n Z_1^T\t^{-1}+
\n Z_1\s Z_1^T\t^{-1}Z_2\n Z_2^T\t^{-1}+\right.
\nonumber
\vspace{-9mm}
\ee
\be
\left.\n Z_2Z_2^T\t^{-1}Z_1\s\n Z_1^T\t^{-1}+
\n Z_2\left(I_n+Z_2^T\t^{-1}Z_2\right)\n Z_2^T\t^{-1}\right],
\ee
or, equivalently,
\be
\S_{matt}=\int d^3x\mid g\mid^{\frac{1}{2}}
Tr\left(\t^{-1}\n Z_kY_{kl}\n Z^T_l\right),
\ee
where the symmetric block--matrix reads
\be
Y_{kl}=
\left(
\ba{cc}
\s+\s Z_1^T\t^{-1}Z_1\s & \s Z_1^T\t^{-1}Z_2 \cr
Z_2^T\t^{-1}Z_1\s & I_n+Z_2^T\t^{-1}Z_2
\ea
\right),
\nonumber 
\ee
$\t=\s-Z_1\s Z_1^T-Z_2Z_2^T$, $I_n$ is the unit matrix of
dimension $n$ and $k,l=1,2$.
The corresponding equations of motion for the matrix potentials 
$Z_1$ and $\Z_2$ are
\be
\n^2Z_1+2\left(\n Z_1\s Z_1^T+\n Z_2Z_2^T\right)\t^{-1}\n Z_1=0
\nonumber \\
\n^2Z_2+2\left(\n Z_1\s Z_1^T+\n Z_2Z_2^T\right)\t^{-1}\n Z_2=0
\ee
This parametrization of the effective theory is a generalization of the 
K\"ahler $\sigma$--model representation of the stationary EM theory
\cite{mazur} in terms of a pair of real matrix potentials instead of 
complex functions. An important feature of this action is its
evident invariance under the transformation 
\be
Z_2\rightarrow Z_1\tau 
\ee
if the rectangular matrix $\tau$ satisfies the following conditions
\be
\tau\tau^T=\s, \qquad \qquad \tau^T\s\tau=I_n. 
\ee

This symmetry mixes the gravitational and 
matter degrees of freedom of the theory. It recalls the Bonnor transformation 
of the Einstein--Maxwell theory \cite{b}--\cite{f}, but in the heterotic string
realm. 

In the particular case when $\tau$ is a square matrix, only the first 
restriction is sufficient to ensure the map (13). This symmetry enables us
to construct new classes of solutions for the whole theory on the basis of
solutions for $Z_1$ or $Z_2$ making use of a simple solution generating procedure. 

In principle, one can study the effective action under investigation formulated
in terms of other dynamical variables in which the group $O(d+1,d+n+1)$ acts linearly 
on the coset space $O(d+1,d+n+1)/O[(d+1)\times (d+n+1)$. In order to achieve this
aim, let us introduce the $O(d+1,d+n+1)$--matrix vector
${\bf W}=\left(W_1,W_2,W_3\right)\ne 0$ with components
defined by the relations
\be
Z_1\equiv (W_2)^{-1}W_1, \quad Z_2\equiv (W_2)^{-1}W_3,
\ee
where $W_1$ and $W_2$ are $(d+1)\times (d+1)$--matrices and the dimension
of $W_3$ is $(d+1)\times n$. Let us define as well the
$O(d+1,d+n+1)$--invariant scalar product in the space of vectors
${\bf W}$
\be
\left({\bf W},{\bf W}^T\right)\equiv \left(W_1,W_2,W_3\right)
\tilde\L\left(W_1,W_2,W_3\right)^T=
-W_1\s{W_1}^T+W_2\s{W_2}^T-W_3{W_3}^T,
\ee
where the matrix $\tilde\L$ determines the indefinite signature 
$\tilde\L=diag(-\s,\s,-I_n)$ of the vector space.

In terms of the introduced vector our action adopts the form
\be
\S=-\int d^3x\mid g\mid^{\frac{1}{2}}
Tr\left\{R+\left({\bf W},{\bf W}^T\right)^{-1}
\left[\left(\n{\bf W},\n{\bf W}^T\right)-\right.\right.
\nonumber
\vspace{-9mm}
\ee
\be
\left.\left.\left(\n{\bf W},{\bf W}^T\right)
\left({\bf W},{\bf W}^T\right)^{-1}\left({\bf W},\n{\bf W}^T\right)
\right]\right\},
\ee
the corresponding equations of motion are
\be
R_{\mu\nu}=
-Tr\left\{\left({\bf W},{\bf W}^T\right)^{-1}
\left[\left(\n_{\mu}{\bf W},\n_{\nu}{\bf W}^T\right)-
\left(\n_{\mu}{\bf W},{\bf W}^T\right)
\left({\bf W},{\bf W}^T\right)^{-1}\left({\bf W},\n_{\nu}{\bf W}^T\right)
\right]\right\},
\nonumber
\ee
\be
\n^2{\bf W}-2\left({\bf W},\n{\bf W}^T\right)
\left({\bf W},{\bf W}^T\right)^{-1}\n{\bf W}=0,
\ee
which is nothing else that a matrix vector generalization of the Ernst 
equation for ${\bf W}$. 

This new dynamical system is related to the original one in the following 
sense: any solution of the field equations (18) can be translated into a
solution of the equations of motion for the original theory using the 
algebraic relations (15).
This formulation of the theory and its equation of motion 
is explicitly $O(d+1,d+n+1)$--invariant and is a direct generalization 
of the representation given in \cite{mazur} and \cite{kinn}
in the framework of the stationary EM theory. The realization of the
linear action of the $O(d+1,d+n+1)$ symmetry group on the coset space
$O(d+1,d+n+1)/O[(d+1)\times (d+n+1])$ is reached by means of the matrix
transformation
\be
{\bf W'}=U{\bf W}
\ee
where the matrix $U$ satisfies the following condition
\be
U\tilde\L U^T=\tilde\L, 
\ee
i.e., $U$ belongs to the $O(d+1,d+n+1)$ symmetry group.

In the simplest case when the matrix vector ${\bf W}$ has the form
\be
{\bf W}=\Psi^{-1}{\bf K}, 
\ee
where the $\Psi$ is a real scalar function and ${\bf K}$ is a constant 
matrix vector, the equation of motion (18) reduces to the Laplace 
equation for the function $\Psi$
\be
\n^2\Psi=0.
\ee
The general solution of this equation is well known and in the simplest case
we can consider the spherically symmetric solution
\be
\Psi=1-\frac{2m}{r}.
\ee 
%%%%%%%%%%%%%%%%%%%%%%%%%%%%%%%%%%%%%%%%%%%%%%%%%%%%%%%%%%%%%%%%%%%%%%%%%%%%%%%%%
\section{Conclusion and discussion}
%%%%%%%%%%%%%%%%%%%%%%%%%%%%%%%%%%%%%%%%%%%%%%%%%%%%%%%%%%%%%%%%%%%%%%%%%%%
%%%%%%%%%%%%%%%%%%%%%%%%%%%%%%%%%%%%%%%%%%%%%%%%%%%%%%%%%%%%%%%%%%%%%%%%%%%%%%%%%
We have presented a couple of symmetric formulations of the toroidally compactified
stationary heterotic string theory. The first representation is written in terms 
of a pair of matrix potentials $Z_1$ and $Z_2$ that enter the effective action in a 
completely symmetric way, a fact that allows us to apply a simple 
solution--generating procedure on the basis of either $Z_1$ or $Z_2$. The second 
parametrization is expressed in terms of a matrix vector ${\bf W}$ which linearizes 
the action of the $O(d+n+1)$ symmetry group on the coset space 
$O(d+1,d+n+1)/[O(d+1)\times O(d+n+1)]$ and can be exploted for generating new 
solutions on the basis of known ones. Under the asumption of a simple ansatz
for the matrix vector ${\bf W}$, its equation of motion reduces to the Laplace 
equation for a real scalar function giving rise to a family of extremal solutions.
 
As a further development of this formalism we would like to address the application 
of solution--generating techniques in the framework of both formulations ($Z_1,Z_2$
and ${\bf W}$) in order to obtain and study more complicated field configurations. 
Another interesting issue is the investigation of the full symmetry  
group of the theory expressed in terms of the matrix vector ${\bf W}$ since it
introduces one more matrix dynamical variable in the formalism.

%%%%%%%%%%%%%%%%%%%%%%%%%%%%%%%%%%%%%%%%%%%%%%%%%%%%%%%%%%%%%%%%%%%%%%%%%%%%%%
%%%%%%%%%%%%%%%%%%%%%%%%%%%%%%%%%%%%%%%%%%%%%%%%%%%%%%%%%%%%%%%%%%%%%%%%%%%%%%
\section{Acknowledgements}
The authors thank Dr. O. Kechkin for helpfull discussions.
This work was supported by grants CIC-UMSNH-4.18 and CONACYT-J34245-E.
AHA is really grateful to S. Kousidou for encouraging him during the 
performance of this research.  

%%%%%%%%%%%%%%%%%%%%%%%%%%%%%%%%%%%%%%%%%%%%%%%%%%%%%%%%%%%%%%%%%%%%%%%%%%%%%%%
%%%%%%%%%%%%%%%%%%%%%%%%%%%%%%%%%%%%%%%%%%%%%%%%%%%%%%%%%%%%%%%%%%%%%%%%%%%%%%%
    
\end{document}